\documentclass[journal=jpclcd,manuscript=article, layout=twocolumn]{achemso}

\usepackage{bbm}
\usepackage{multirow}
\usepackage{float}
\usepackage{graphicx} 
\usepackage[cmyk,dvipsnames]{xcolor}
\usepackage{braket}

\author{Alejandro Peña-Torres}
\affiliation[University of Iceland]
{Science Institute and Faculty of Physical Sciences, U. of Iceland, 107 Reykjavík, Iceland}
\author{Michail Stamatakis}
\affiliation[University of Oxford]
{Departmentof Chemistry, Inorganic Chemistry Lab, University of Oxford, Oxford OX1 3QR, U.K.}
\author{Hannes Jónsson}
\affiliation[University of Iceland]
{Science Institute and Faculty of Physical Sciences, U. of Iceland, 107 Reykjavík, Iceland}
\email{hj@hi.is}

\title{Dimerization of Ag adatoms on Si(100) at surprisingly low temperature due to adatom migration on top of Si-dimer rows}

\begin{document}

\begin{tocentry}
\includegraphics[width = \linewidth]{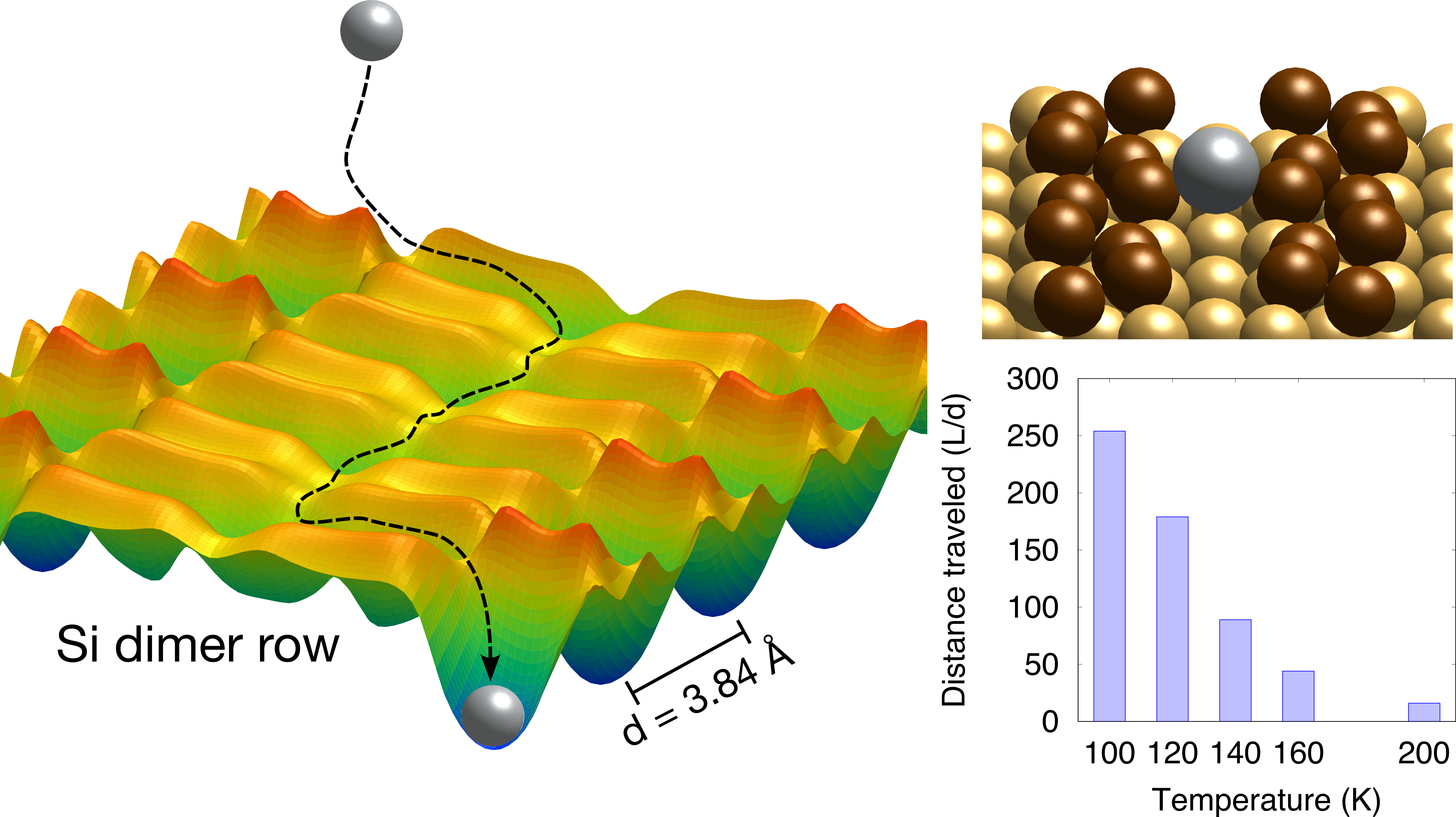}

\end{tocentry}

\begin{abstract}
The puzzling experimental observations of Ag addimer formation on the Si(100) surface upon vapor deposition at low temperature, even as low as 140 K, reported by Huang et al. [Phys. Chem. Chem. Phys. 2021, 23, 4161], is explained by facile thermal migration on top of the Si dimer rows, while diffusion is inactive once an Ag adatom lands in one of the optimal binding sites in between dimer rows. The previously hypothesized transient mobility due to the hot spot formed as an Ag atom lands on the Si surface is found not to contribute significantly to mobility and dimer formation. The experimental conditions are simulated by a combination of classical dynamics calculations of the deposition events, systematic searches of saddle points representing transition states for thermally activated events and kinetic Monte Carlo simulations of long timescale evolution of the system over a temperature range from 100 K to 300 K. While the calculated energy barriers for diffusion parallel and perpendicular to the Si dimer rows are found to be nearly equal, indicating isotropic diffusion, the simulations show highly anisotropic diffusion, as the optimal migration mechanism involves multiple hops of Ag adatoms on top of Si dimer rows. Impinging Ag atoms are found to have a 90\% chance of landing on top of a dimer row and while the hot spot created cools down too fast for transient mobility to play an important role, the energy barrier for thermally activated hops along the dimer row is low enough for migration to be active even at 100 K. A migrating Ag adatom can dimerise with another Ag adatom sitting at a stable binding site in between rows or another adatom on top of the same row. The simulations for deposition at 140 K show that most of the deposited Ag atoms end up forming dimers. 
\end{abstract}




Metallic thin films and nanostructures on semiconductor surfaces are of great importance for various devices, 
with silicon substrates receiving particular attention due to their prominence in electronics.
Silver deposited on Si(100) surfaces has become one of the most frequently studied metal/semiconductor systems and a great deal of work has been done to study nanostructure formation and overlayer growth. Various experimental techniques have been applied to this system such as coaxial impact collision ion scattering spectroscopy~\cite{Kim1998,Cho1999}, photoemission spectroscopy~\cite{Matsuda1999, Matsuda2001}, scanning tunneling microscopy (STM)~\cite{Hashizume1990,Winau1994,Kocan2006,Takeuchi2011} and micro-calorimetry~\cite{Campbell2001}.

There has been ongoing discussion about the diffusion mechanism of Ag adatoms and Ag aggregates formation on the dimer row reconstructed Si(100) surface. Huang \textit{et al.} attributed the experimental observation of Ag agreggates at low temperatures to hot-atom motion upon initial collision of single Ag atoms on the Si surface, indicating that thermal diffusion is negligible at such low values of temperature.~\cite{Huang2021} However, there has been no further experimental or theoretical evidence to support this claim.

Regarding the single Ag atom diffusion, early work by Zhou \textit{et al.}~\cite{Zhou1993} concluded that the migration is anisotropic with diffusion favored perpendicular to the dimer rows and with an activation energy of only 0.11 eV. On the other hand, Okon \textit{et al.}~\cite{Okon1997} reported preferred migration parallel to the dimer rows. More recently, extensive studies combining cryogenic STM experiments and density functional theory (DFT) calculations have been carried out
~\cite{Huang2018, Singh2019, Huang2021, Huang2022}. From the calculated energy landscape, Huang \textit{et al.}~\cite{Huang2022} conclude that the adatom diffusion is isotropic on the c(4$\times$2) reconstructed surface
with an energy barrier of 0.51 eV. This agrees with earlier theoretical calculations by Kong \textit{et al.}~\cite{Kong2003}, who also reported an isotropic barrier of 0.5 eV for the adatom migration on the p(2$\times$2) reconstructed the Si(100) surface.

Here, we present theoretical calculations for the deposition, diffusion and possible dimerization of Ag atoms on the dimer row reconstructed Si(100) surface. 
Molecular dynamics simulations of a single Ag atom deposition on the Si surface using a Tersoff-type potential parametrized for the Ag/Si interaction\cite{Butler_2011} show that the preferred landing sites lie on top of the Si dimer rows. And that the energy of the deposited atom is dissipated after 1-2 ps, making transient mobility upon initial impingement unlikely at low temperatures. 
To study the long time scale dynamics of a single Ag adatom diffusing on the Si(100) surface at low temperatures, we use the adaptive kinetic Monte Carlo (AKMC) method. Our simulations performed at 140~K show that the adatom performs around 60 hops between binding sites on top of the dimer rows before falling down to the optimal binding site in between the dimer rows.
To corroborate our results, we use the nudge elastic band (NEB) method to calculate the migration paths with atomic forces from density functional theory (DFT) calculations within the PBEsol functional approximation, and subsequent kinetic Monte Carlo (KMC) simulations of the diffusion over a temperature range from 100 K to 300~K, we find that the diffusion is anisotropic with faster diffusion parallel to the dimer rows.
While the energy barrier obtained from the energy landscape indeed indicates isotropic diffusion, the anisotropy results from fast migration along the top of a dimer row in between visits to optimal binding sites in between dimer rows.
From the temperature dependence of the diffusion coefficient obtained from the KMC simulation, an effective activation energy for parallel diffusion is 20\% lower than what would be expected from the highest energy barrier that needs to be overcome.
Additionally, we present KMC simulations at 140~K where Ag dimers can form when two adatoms meet during diffusion.
The preferred migration on top of the dimer rows turns out to play a key role in the fast formation of Ag addimers.
These results can help explain why single Ag adatoms are not observed in STM experiments, even at low temperature where thermal migration is expected to be inactive.

\begin{figure}[!b]
    \centering
    \includegraphics[width=0.6\columnwidth]{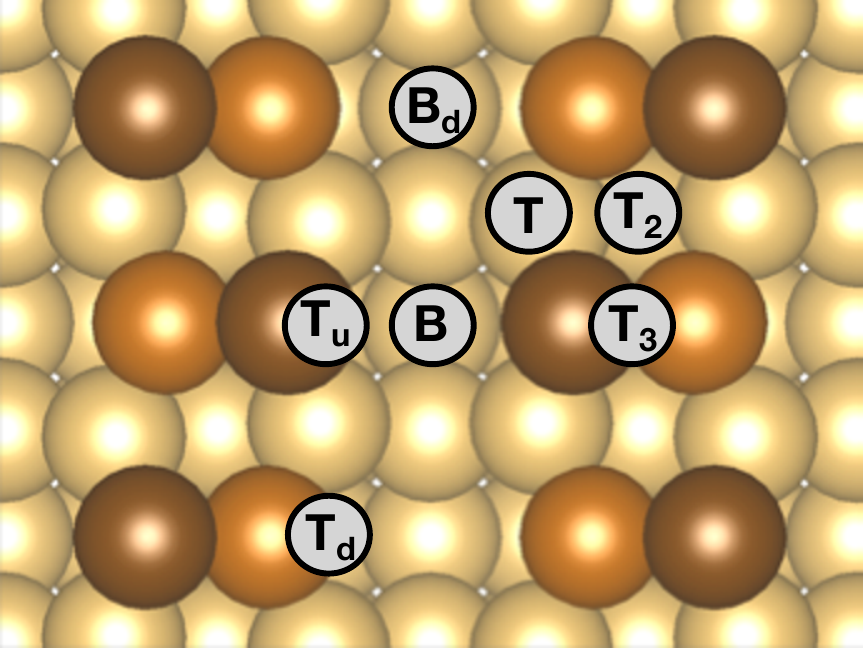}
    \caption{
    Top view of the clean Si(100) displaying a c(4$\times$2) rearrangement. The buckled dimers are highlighted with a brown (orange) color for the upper (lower) Si atoms. 
    The binding sites found for the Ag adatom, in between dimer rows (B, B$_d$) and on top of dimer rows (T, T$_2$, T$_3$, T$_u$ and T$_d$) are marked on the surface. }
    \label{fig:SiSlab_adsSites}
\end{figure}


To assess the initial impingement and preferred landing sites of the Ag atoms on the Si surface, we carried out 1000 molecular dynamics simulations using a Tersoff-type interatomic potential parametrized to describe the Ag/Si interaction.~\cite{Butler_2011} Each simulation started by positioning the Ag atom 7~\AA~from the Si surface with random $x$ and $y$ coordinates. The initial velocity corresponds to the average velocity from a Maxwell-Boltzmann distribution at room temperature. Remarkably, in 91\% of these simulations, the Ag atom stabilized atop the dimer rows, despite the optimal binding site being located between them (see Fig. S3). The kinetic energy of the impinging atom is dissipated within 1-2~ps. Furthermore, the average displacement on the surface from the moment the atom landed --defined as when the velocity in the $z$ direction becomes positive-- until the kinetic energy dissipated was 1.2~\AA. Given that the distance between adjacent binding sites atop the dimer rows is approximately 2~\AA, this suggests that roughly half the time an atom lands, it performs one hop due to transient mobility from impact. Contrary to previous reports suggesting diffusion driven by hot-atom motion~\cite{Huang2021}, our results suggest that the formation of silver aggregates is likely due to alternative diffusion mechanisms rather than transient mobility.

The AKMC simulations at 140~K starting with the Ag atom adsorbed on top of the dimer rows result in the atom performing 59 hops between different adsorption sites before falling down to the optimal site in between the dimer rows. The simulated time until this process occurs is 12800~s ($\approx$ 213 min). It is worth noting that the Tersoff potential used in these simulations does not reproduce the buckling pattern of the Si(100) reconstructed surface. However, the energy barriers between adsorption sites and overall energy landscape compare well with our DFT calculations, as discussed below.

In order to obtain the optimal structure of the reconstructed Si(100) surface, 
a minimization is carried out starting from the unreconstructed surface where there are two dangling bonds for each surface Si atom. 
After optimization of the atomic coordinates, the reconstructed Si(100)
surface is found to have the c(4$\times$2) structure where the tilting of the dimers in adjacent Si dimer rows follows opposite patterns
(see Fig.~\ref{fig:SiSlab_adsSites}).
This reconstruction is found to be slightly more stable than the p(2$\times$2) structure. 
The c(4$\times$2) structure has been found in STM experiments to be the predominant configuration of the surface \cite{Wolkow_1992}.
It has also been deduced from classical molecular dynamics simulations \cite{Zhao_2014}.  
The resulting surface has a dimer bond length of 2.34~\AA~and a buckling angle of 19.8$^\circ$. These values are in good agreement with data obtained from low energy electron diffraction (LEED) experiments giving values of 2.24~\AA~and 19.2$^\circ$ for the dimer bond length and buckling angle, respectively \cite{Over_1997}. 

\begin{figure}[t]
    \centering
    \includegraphics[width=0.45\textwidth]{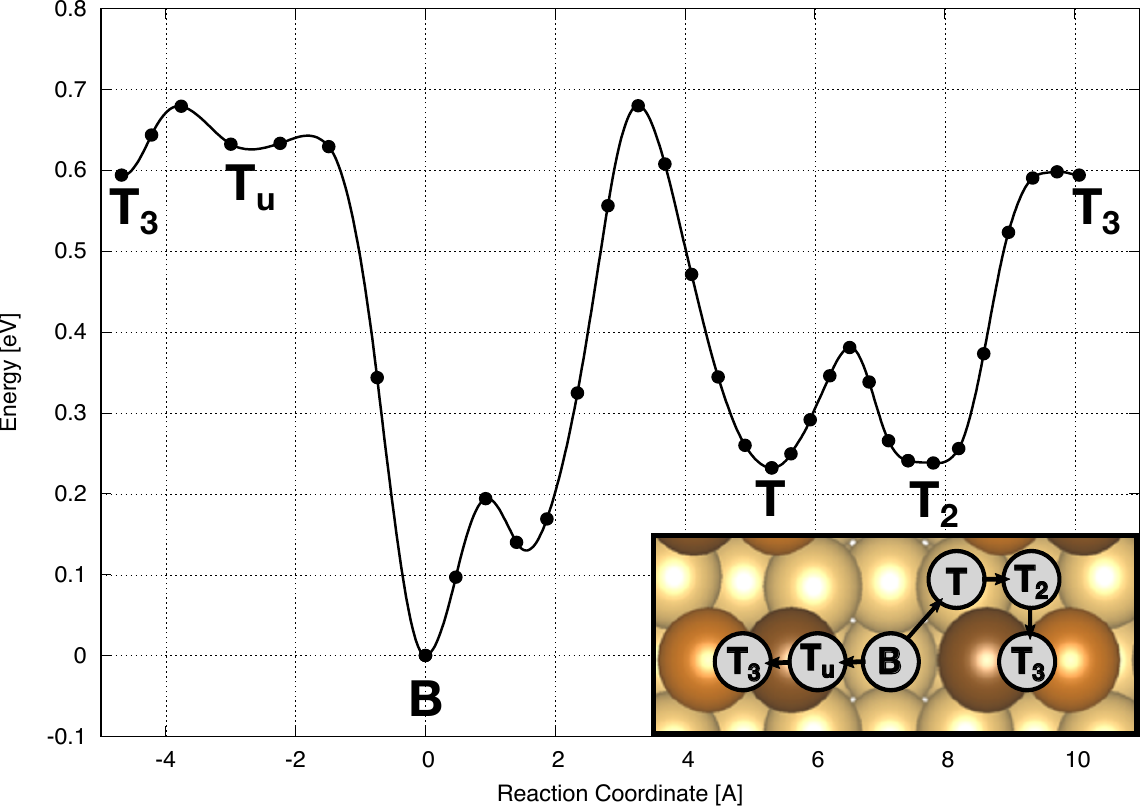}
    \caption{
    Energy along calculated minimum energy paths between the various adsorption sites for a Ag adatom on the dimer row 
    reconstructed Si(100) surface. 
    The filled disks correspond to images in the CI-NEB calculations of the paths.
    The inset illustrates the two possible paths that the Ag adatom can take to hop from a B site to a T$_3$ site.
    }
    \label{fig:MEPs}
\end{figure}

To identify the adsorption sites of the Ag adatom on the surface, the adatom is placed at various initial locations above the surface followed by a minimization of the energy with respect to the coordinates of the moving atoms.
After a local energy minimum has been reached, the binding energy, $E_{b}$, is evaluated as the difference between the energy of the silicon slab with and without the Ag adatom, plus the energy of an isolated Ag atom.
In this as well as previous studies, the optimal binding site is found to be in between dimer rows, in a location where the 
nearest Si atoms are in the upper position of the tilt. This is referred to as the B site.
Several other sites are also identified as local minima on the energy landscape, as shown in Fig.~\ref{fig:SiSlab_adsSites} and listed
in Table~S1.
The second and third most stable sites are the asymmetric site on the edge of a dimer row, labeled as T, 
and the site on top of a dimer row, labeled as T$_2$.
In addition to the listed sites, another local minima on top of the dimer rows was identified, where the Si dimer is dissociated and the Ag adatom is located in between the Si atoms, with a corresponding binding energy of 1.64~eV. 

The results of the PBEsol calculations presented here of the binding energy at the various sites are in good correspondence 
with the published results based on the PBE functional approximation in Ref.~\cite{Huang2022}. The previous results based on
the more approximate LDA functional gave larger binding energy as would be expected. In general, the LDA functional gives 
larger overestimation of binding energy than the generalized gradient approximation functionals such as PBE and PBEsol (See Table S1).

The NEB calculations reveal two possible paths for the Ag adatom to hop on top of a Si-dimer row, which are shown in Fig.~\ref{fig:MEPs}.
Generating an initial IDPP path for going from one optimal binding site, B, to an adjacent B site 
and then carrying out the CI-NEB optimization reveals that the MEP goes through a local minimum at the edge of a dimer row in between two Si-dimers, the T site (see Fig.~\ref{fig:SiSlab_adsSites}).
The activation energy for the B$\rightarrow$T hop is 0.68~eV, while the opposite process, to jump back to the same or an adjacent B site, has an activation energy of 0.45~eV.
The most direct diffusion path parallel to the dimer rows is therefore B$\rightarrow$T$\rightarrow$B$\rightarrow$T\dots .

However, once the Ag adatom reaches a T site it can also hop to a T$_2$ in the center of the dimer row in between two Si-dimers. 
This T$\rightarrow$T$_2$ hop has an activation energy of 0.15~eV.
Subsequently, the adatom can hop along the Si dimer row by going to an adjacent T$_2$ site by overcoming an energy barrier of 0.36~eV. 
This process involves passing through a configuration T$_3$ where the Ag adatom is located on top of a Si-dimer. 
It is barely a minimum on the energy surface as there is only a barrier of  
0.005~eV in order to hop back to a T$_2$ site on either side.
An alternative path for diffusion parallel to the dimer rows therefore involves the following sequence of hops:
B$\rightarrow$T$\rightarrow$T$_2$$\rightarrow$T$_3$$\rightarrow$T$_2$\dots
There, the adatom makes a (potentially large) number of hops along the top of the dimer row before sliding down to an optimal B site again.
The highest energy barrier of this second migration path is the same as for the first one since it involves the same initial
B$\rightarrow$T hop.

The path for migration perpendicular to the dimer rows also involves visiting the barely existing T$_3$ site and from there
another shallow minimum T$_u$ before hopping to an optiml B site on the other site of the dimer row.
The overall energy barrier for perpendicular diffusion turns out to be the same as for parallel diffusion, 0.68~eV. 
From the energy barriers, calculated from the highest point on the MEP minus the initial state energy, diffusion parallel and
perpendicular to the dimer rows have a difference of less than 0.01~eV in their activation energy.

The Ag adatom located at a B site has six ways to leave the site, four of them leading to a T site, and two of them leading  to a T$_3$ site passing through T$_u$.
The fact that the energy barrier is practically the same for both parallel and perpendicular diffusion could imply that the diffusion on the surface is isotropic, as has been suggested previously~\cite{Huang2022,Kong2003}.
Nevertheless, once the Ag adatom lands on the T site, it needs slightly less energy (0.09~eV) to diffuse along the Si dimer rows than to slide down to a B site.
This difference in energy barriers can lead to a preferred parallel diffusion, particularly at low temperatures.


To investigate the effect of the different diffusion paths, KMC simulations have been carried out for Ag adatom diffusion over a range of temperatures. The elementary transitions used as input for the KMC simulations correspond to the following reversible processes:
(i) B$\longleftrightarrow$T; (ii) B$\longleftrightarrow$T$_3$; (iii) T$\longleftrightarrow$T$_2$; (iv) T$_2\longleftrightarrow$T$_3$.
The shallow T$_u$ minimum is not included.
The energy barrier for each elementary hop as well as the binding energy for each site is obtained from the DFT/PBEsol calculations as explained above but the pre-exponential factor in the rate constant expression is assumed to be the same, equal to
10$^{12}$ s$^{-1}$.
The lattice used in the KMC simulations includes the B, T, T$_2$ and T$_3$ sites as shown in the inset of Fig.~\ref{fig:MEPs}.

\begin{figure}[!h]
    \centering
    \includegraphics[width=0.45\textwidth]{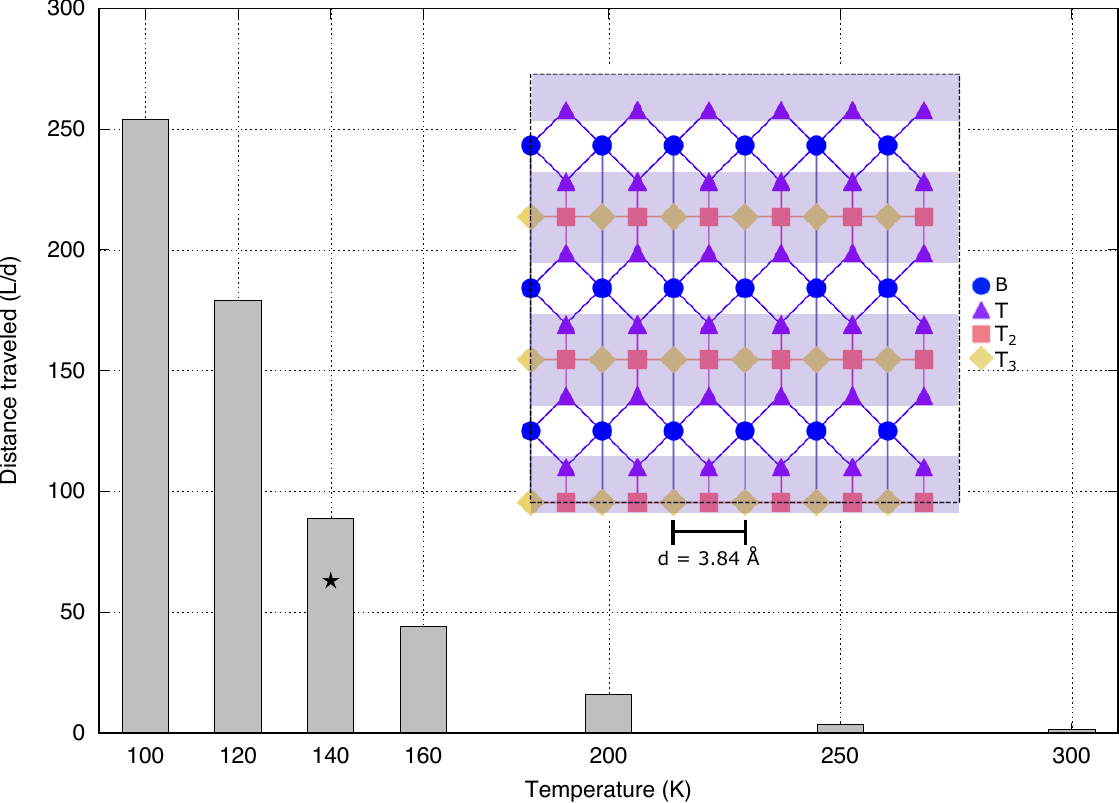}
    \caption{
Average number of Ag adatom hops along the Si dimer rows in between visits to the optimal B sites obtained from the KMC
simulations at various temperature values. 
The inset shows the binding sites lattice used in the KMC simulations, with the Si dimer rows shaded. The star marker at 140~K indicates the number of hops obtained in the AKMC simulations using the Tersoff-type potential.
    }
    \label{fig:kmclattice}
\end{figure}

The KMC simulations are carried out for seven temperature values in the 100 - 300 K range.
For each temperature value, 10 separate simulations are carried out with different random number seeds.
Fig.~\ref{fig:kmclattice} shows the average number of hops performed by the Ag adatom along a Si dimer row in between visits to a B site. 
An illustration of the KMC lattice used in the simulations is shown as an inset in Fig.~\ref{fig:kmclattice}, where the Si dimer rows are shaded for illustrative purposes.

\begin{figure*}[!h]
    \centering
    \includegraphics[width=0.95\textwidth]{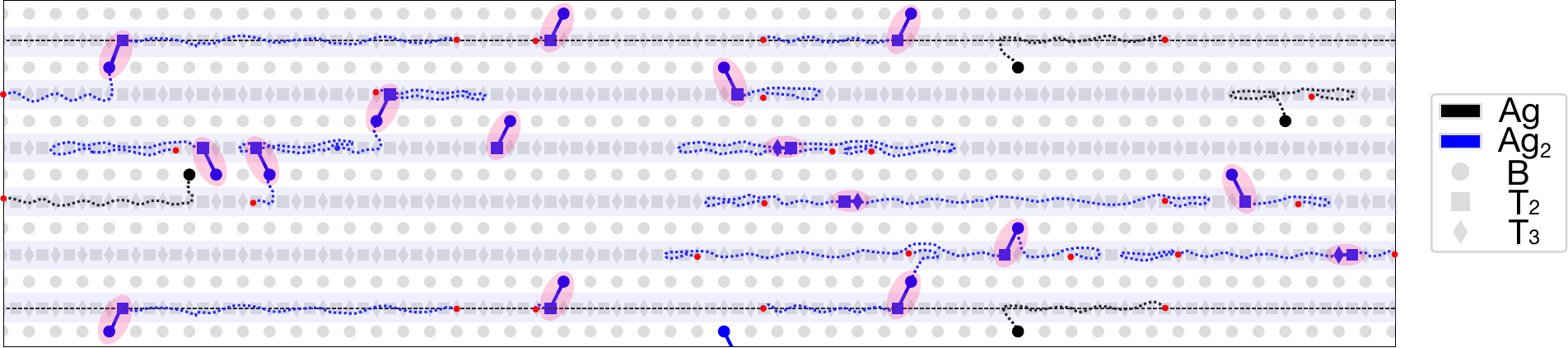}
    \caption{
Snapshot of a simulation at t = 1050 s. The dotted lines represent the boundaries of the lattice and the initial position of the preadsorbed Ag adatoms that hop along the surface during the simulation are marked in red. The Si dimer rows are shaded for illustrative purposes. The trajectory followed by the adatoms is marked in blue (black) dotted lines for the atoms that ended up forming a dimer (isolated) at the end of the simulation. The formed dimers are also highlighted for clarity.
    }
    \label{fig:DimersKMC}
\end{figure*}

The results show that the average number of hops along a dimer row in between visits to B sites increases exponentially as the temperature is lowered.
At low temperature, the Ag adatom is more likely to hop along a dimer row after leaving a B site, instead of falling down to an adjacent B site.
This effect arises from the difference in energy barriers between sliding along a dimer row from a T$_2$ site (0.36~eV) and jumping back down to a B site (0.45~eV). This difference in the energy barriers of 0.09~eV has a significant effect at low temperature.
At room temperature, however, the difference in the energy barriers is less important, and the Ag adatom diffusion becomes almost isotropic.
Results from the AKMC simulation of the Ag adatom diffusion at 140 K show that the number of hops before falling down to a B site is 59 (star marker in Fig.~\ref{fig:kmclattice}). This decrease with respect to the KMC simulations is to be expected given that the energy barriers for diffusing on top of the dimer rows are larger in the Tersoff-type potential than in the DFT/PBEsol case (See Fig. S4 for a comparison of the energy landscapes). 

From the KMC simulations, the average distance traveled by the Ag adatom in between visits to B sites can be obtained.
The diffusion coefficient ($D$) can also be obtained assuming a 1D random walk between B sites as $D(T)=\frac{<L^2>}{2\tau}$,
where $L$ is the average distance travelled and $\tau$ is the average time spent at a B site before hopping on top of a dimer row. 
From the slope of the linear fit over this temperature range an effective diffusion activation energy of 0.55~eV is obtained (See Fig. S1).
This means that, even though the energy barrier that needs to be overcome for both parallel and perpendicular diffusion is the same, the effective activation energy for parallel diffusion ends up being ca. 20\% lower than the energy barrier 
because of the repeated hops on top of the dimer rows once the adatom leaves a B site.
The deviation from a linear fit arises from the difference in diffusion mechanisms at low temperature.
At temperature above 300~K the rate of parallel and perpendicular diffusion processes will be the nearly same, resulting in isotropic diffusion.


The reduction of the effective activation energy for parallel diffusion due to the fast migration on top of dimer rows can help explain puzzling experimental STM observation where no isolated Ag adatoms are found, only dimers,
even after deposition at temperature as low as 100 K~\cite{Huang2021}. 
To investigate this, we performed KMC simulations at 140 K including the possibility of Ag dimers formation when two single adatoms end up in adjacent sites during diffusion. 
Here, the T and T$_2$ sites were merged,
and the dimers were allowed to be formed in the B--(T+T$_2$) and (T+T$_2$)--T$_3$ configurations.

The probability for an atom to land on top of the dimer rows was taken as four times greater than that of landing in between rows, given that we have four possible adsorption sites on top of the dimer rows whereas only one in between (T, T$_2$, T$_3$, T$_u$ vs. B). This agrees with our findings from the Ag deposition simulations where the single Ag atom ended up landing on top of the dimer rows 91\% of the times.

Fig.~\ref{fig:DimersKMC} corresponds to a snapshot of the lattice at t = 1050 s where the trajectory followed by the adatoms is marked in blue dotted lines for the atoms that ended up forming a dimer in the simulation (highlighted) and black dotted lines for the atoms that reached a B site without forming a dimer. The initial position for the atoms that diffused during the simulation is marked in red. From the simulation, we can see that most of the atoms that started on top of the dimer rows diffused along it before either forming a dimer with another Ag atom, or simply hoping down to a B site.

We also carried out a simulation using the same initial conditions and artificially increasing the energy barrier for the T$_2\longrightarrow$~T$_3$ process to 1~eV so as to prevent diffusion along the dimer rows. As expected, this resulted in most of the atoms not forming a dimer at the end of the simulation (See Fig. S2). From these results, we observe that the migration along the dimer rows allows single adatoms to either form a dimer with an adatom sitting in a B site or to form it with another adatom skidding on top of the dimer rows. When the energy barrier for sliding along the dimer rows is increased, the preadsorbed adatoms hop down to a B site, where they are unable to keep diffusing at this timescale (the time to leave a B site at 140 K is ca. 5x10$^{11}$ s). 

Previous studies of other types of adatoms on the dimer row reconstructed Si(100) surface have revealed  similar diffusion mechanisms, namely for a Si adatom~\cite{Smith_Jonsson_PRL1996} and Au adatom~\cite{PenaTorres2022}. Also there, the path for migration along the top of Si dimer rows turns out to be faster than a direct transition between optimal  sites in between
dimer rows. This explained the experimental observation that Si addimers form primarily on top of dimer rows even though the energetically most stable configuration is in between dimer rows\cite{Smith_Jonsson_PRL1996}. 




In conclusion, our study methodically unraveled the deposition and diffusion mechanisms of a silver atom on the Si(100) surface. Initially, we employed a Tersoff potential, specifically parameterized for the Ag/Si interaction, to simulate the deposition process through molecular dynamics. This approach was complemented by adaptive kinetic Monte Carlo (AKMC) simulations, which provided detailed insights into the diffusion dynamics of Ag adatoms along the dimer rows, before they settled into the optimal binding sites between these rows. Subsequently, we determined the energy barriers between these local minima using the climbing image nudged elastic band (CI-NEB) method with energy and forces obtained from DFT/PBEsol. Finally, we performed KMC simulations based on the DFT energy landscape to further investigate the diffusion in the 100 - 300 K temperature range and the possible dimerization of Ag atoms.

From the molecular dynamics simulations of Ag deposition on the surface, we notice that the formation of Ag aggregates is not driven by hot-atom motion by initial impact, since the kinetic energy of the impinging atom is dissipated within 1-2 ps. Instead, it stems from structural guidance by the Si dimer rows, as seen from the AKMC simulation.

Consistent with previous research, the most stable adatom site is between the dimer rows, stabilized by two Si atoms. From this position, an Ag adatom has six possible pathways, each requiring an activation energy of 0.68~eV.

Our results emphasize that adatom migration atop Si dimer rows, particularly at low temperatures, is predominantly parallel. Adatoms move along the dimer rows before returning to optimal binding sites between rows. The effective activation energy for this diffusion, determined from temperature-dependent KMC simulations, is about 0.55 eV. This is approximately 20\% lower than the energy barrier for direct diffusion.

This preferred migration parallel to the dimer rows at low temperature 
leads to anisotropic diffusion while the energy barriers alone indicate isotropic diffusion, as has been noted previously. Our simulations show that even at low temperatures, where thermal energy is minimal, adatoms can rapidly form aggregates.


\section{Methods}

The AKMC method as implemented in the EON~\cite{EON} package was used to simulate the diffusion of the single Ag adatom on the Si(100) surface. The energy and the forces at each step were evaluated with a Tersoff-type interatomic potential~\cite{Tersoff_1986}, parametrized by Butler {\it et al.} for the Si/Ag interaction.~\cite{Butler_2011} The simulation was performed at 140~K. At each state, saddle searches were performed until a confidence of 0.99 was reached. The pre-exponential factor that goes in the Arrhenius equation to calculate the rate constant for each process was set to 10$^{12}$ s$^{-1}$.

The molecular dynamics simulations for the single Ag atom deposition on the Si surface were performed employing a Tersoff-type potential described above. These calculations were carried out using the DL\_POLY package.~\cite{DLPOLY} The initial $x$ and $y$ coordinates of the Ag atom in each of the 1000 simulations were randomly sampled at a distance of $z=7$~\AA~ from the surface, and the initial velocity was chosen as the average velocity from a Maxwell-Boltzmann distribution at room temperature (0.033 eV in this case). The timestep was set to 1 fs.

For the DFT calculations, the dimer row reconstructed Si(100) surface is represented by a periodic 4$\times$4 surface supercell slab with six atomic layers constituted by 96 silicon atoms, where the bottom layer is passivated with 16 hydrogen atoms. 
Atoms in the three upper layers are allowed to relax but the rest of the atoms are kept frozen at the perfect crystal positions. 

The climbing image NEB (CI-NEB) method~\cite{Henkelman_2000a,Henkelman_2000b,Asgeirsson2018b} is used to calculate the minimum energy paths between the binding sites of the Ag adatom on the surface.
The initial paths are generated with the image dependent pair potential (IDPP) method~\cite{Smidstrup_IDPP_2014} with six intermediate images. 
The optimization of atomic coordinates is carried out until the magnitude of the components of the atomic forces 
perpendicular to the path are below 0.01~eV/{\AA}.

The energy and atomic forces of the system are calculated using DFT within the PBEsol functional approximation \cite{Perdew_2008_pbesol}.
The PBEsol functional is chosen because it is adjusted to give good estimates of lattice constants for solids and
turns out to give better estimate of surface energy than the PBE functional.
A plane wave basis set is used with a cutoff energy of 350~eV to represent the valence electrons,
while the effect of the inner electrons is represented with the projector augmented-wave (PAW) method \cite{Bloechl94}.

The lattice constant of the crystal within the PBEsol approximation is found to be 5.43~\AA, reproducing the experimental value.
In the slab calculations, the Brillouin zone is sampled with a uniform mesh of 7$\times$7$\times$1 k-points.
The calculations of local energy minima and NEB calculations of minimum energy paths (MEPs) are carried out with the 
EON software~\cite{EON} using energy and atomic forces obtained from the VASP code~\cite{VASP_1,VASP_2}.

The energy barriers, $E_a$, corresponding to the elementary migration processes of the Ag adatom on the Si surface are obtained from the minimum energy paths as the difference between the maximum energy along the path and the energy of the initial state local minimum.
Using the Arrhenius expression, $k = \nu \exp{(-E_a/k_BT)}$, we estimate the rate constant for each process assuming the pre-exponential factor has a typical value of $\nu$~= 10$^{12}$ s$^{-1}$. 
This information 
is then used as input for KMC simulations of the diffusion of the Ag adatom over a range of temperature using the Zacros software~\cite{Zacros_2011,Nielsen_2013,Zacros_2022}. 
The adsorption or possible desorption processes are not included in the KMC simulations.


\begin{acknowledgement}
This project was funded by European Union’s Horizon 2020 research and innovation programme under Grant Agreement No. 814416 (ReaxPro), and the University of Iceland Research Fund. We thank Keith Butler for sharing the parameter values of the potential energy function representing the Ag/Si interaction. 
The calculations were carried out at the IREI computer facility located at the University of Iceland.
\end{acknowledgement}





\bibliography{refs}

\end{document}